# Optically Controlled Supercapacitors with Semiconductor Embedded Active Carbon Electrodes

H. Grebel

The Center for Energy Efficiency, Resilience and Innovation (CEERI), The Electronic Imaging Center (EIC), The New Jersey Institute of Technology (NJIT), Newark, NJ 07102.
grebel@njit.edu

**Abstract:** Supercapacitors, S-C - capacitors that take advantage of the large capacitance at the interface between an electrode and an electrolyte - have found many short-term energy applications. We concentrate here on optically induced, electrical and thermal effects. The parallel plate cells were made of two transparent electrodes (ITO), each covered with semiconductor-embedded, active carbon (A-C) layer. While A-C appears black, it is not an ideal blackbody absorber that absorbs all spectral light indiscriminately. In addition to relatively flat optical absorption background, A-C exhibits two distinct absorption bands: in the near-IR and in the blue. The first may be attributed to absorption by OH- group and the latter, by scattering, possibly by surface plasmons. Here, optical and thermal effects of sub-micron size SiC particles that are embedded in A-C electrode, are presented. Similarly to nano-Si particles, SiC exhibits blue band absorption, but it is less likely to oxidize. Using Charge-Discharge (CD) experiments, the relative optically related capacitance increase may be as large as ~34% (68% when the illuminated area is taken into account). Capacitance increase was noted as the illuminated samples became hotter. This thermal effect amounts to <20% of the overall relative change using CD experiments. The thermal effect was quite large when the SiC particles were replaced by CdSe/ZnS quantum dots; for the latter, the thermal effect was 35% compared with 10% for the optical effect. When analyzing the optical effect one may consider two processes: ionization of the semiconductor particles and charge displacement under the cell's terminals - a dipole effect. Our model suggests that the capacitance increase is related to an optically induced dipole.

## I. Introduction

Supercapacitors, S-C, [1-6] either symmetric (the two cell's electrodes are of the same type), or asymmetric (made with two types of electrodes) [7-9] are capacitors that take advantage of the capacitance at the interface between an electrode and an electrolyte. They have found many energy applications due to their fast charging and discharging. Applications have also been proposed for the future digital micro-grid [10-11] and optical modulators [12]. Here, we describe carbon based, optically controlled S-C that exhibit electrical double-layer behavior. Our intent is to gain basic understanding of the optical and related thermal effects when incorporating SiC particles in active carbon (A-C) based electrodes. We concentrate on PMMA binder and an aqueous electrolyte (1 M $Na_2SO_4$).

## II. Methods and Experiments

The basic cell is composed of two transparent electrodes. Glass substrates were coated with ca 1 micron indium tin oxide (ITO, sheet resistance Rsqr=5 Ohms, Huanyu) and were facing each other to form a parallel plate capacitor (Fig. 1a). The electrodes were coated with A-C film (produced by General Carbon Company, GCC). The sample was held by two metal clips. Scanning Electron Microscopy (SEM) and Atomic Force Microscopy (AFM) were used to assess the sample surface.

**A-C in a PMMA binder with SiC particles:** Typically, a 100 mg/mL AC produced by GCC with a 10 mg/mL PMMA binder. The 10 mg/mL of SiC particles (Johnson Metthey) were mixed with the other components in toluene and were dried out in an oven at 90 $^oC$ for 30 min (Fig. 1). A typical composite A-C film was ~100 microns.

**Electrolyte:** 1 M of $Na_2SO_4$; hydrophilic nano-filtration polyamide filter (TS80, Sterlitech) used as a separator.

**Electrochemical Measurements:** Measurements were carried out with a Potentiostat/Galvanostat (Metrohm). The samples were illuminated by a 75 W incandescent light bulb situated 30 cm or 62.5 cm above the samples. The light intensity of the entire radiation spectra (from the visible to the IR) was measured with a bolometer and was assessed as 30 mW/cm$^2$ and 3 mW/cm$^2$, respectively. Note that the light intensity at the sample surface did not follow an inverse square law for the distance because the light source is extended not a point source. A calibrated homemade hot plate, which was interfaced with a thermocouple was used for the thermal experiments. A second thermocouple assessed the temperature right at the sample surface.

**Optical transmission measurements:** A computer controlled monochromator (SPEX), which was interfaced with a white light source, a chopper and a Si detector was used to assess the optical transmission of each film when deposited on a glass substrate. The transmission value is defined as the signal obtained with the film on a glass slide divided by the signal obtained with only the glass slide. Each transmission curve was normalized to its peak transmission since the thickness of the heterogeneous film was unknown and we are interested in the spectral trend rather than the actual film loss.

## III. Results and Discussion

### III.a. *Characterizations*

**Imaging via SEM and AFM:** Pictures of the surface morphology are shown in Fig. 1b-e. As can be seen from the pictures, the AC domains is of the order of 15 microns. The dispersion of the SiC particles is quite large and ranges between sub-microns to microns.

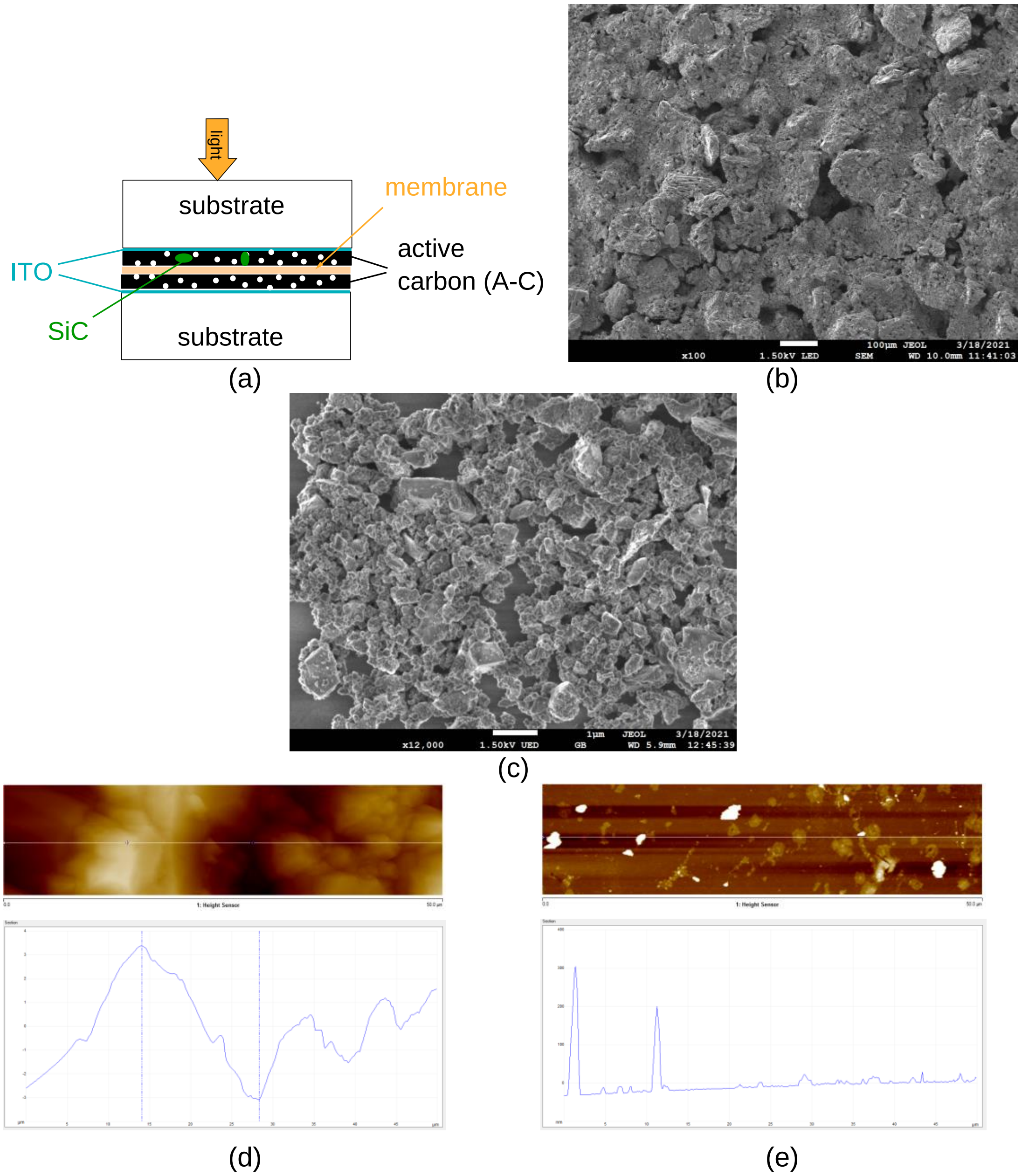

Fig. 1. (a) Schematics of the sample: TS80 is the hydrophilic membrane; ITO: indium-tin-oxide; (b) SEM picture of a films composed of A-C with PMMA binder. The bar is 100 microns. (c) SEM picture of sub-micron SiC particles on glass. The bar is 1 micron. (d) AFM profile of the A-C/PMMA film. (e) AFM profile of SiC particles on glass.

**Optical Transmission:** Optical transmission through the various material components is shown in Fig. 2. Each transmission curve was normalized to its peak transmission. The glass slide signal was referenced to the transmission through air; all other curves were referenced to the signal of their substrate - a glass slide. The transmission of a glass slide is fairly constant throughout the spectral range between 400 to 900 nm (not shown). The transmission of the ITO film on glass is flat throughout the visible with a small absorption near the blue region of 400 nm; PMMA has also a flat transmission in the visible with a transmission coefficient of 0.9 (not shown).

The green/yellow alpha SiC powder absorbs heavily in the blue as are n-Si and A-C (see Fig. 2a). The response of only A-C with a PMMA binder to wet and dry conditions is shown in Fig. 2b. The sample was first soaked in 0.1 M $Na_2SO_4$ for a couple of hours and its optical absorption was assessed. Then the sample was dried up in an oven for three days at 90 $^oC$. One may observed that the general absorption pattern was retained. There are three major absorption peaks (transmission dips) at 430, 670 and 840 nm, respectively. The latter is attributed to $OH^-$ group because it is accentuated by the presence of water. The middle is attributed to the glass substrate. The more interesting one is at 430 nm.

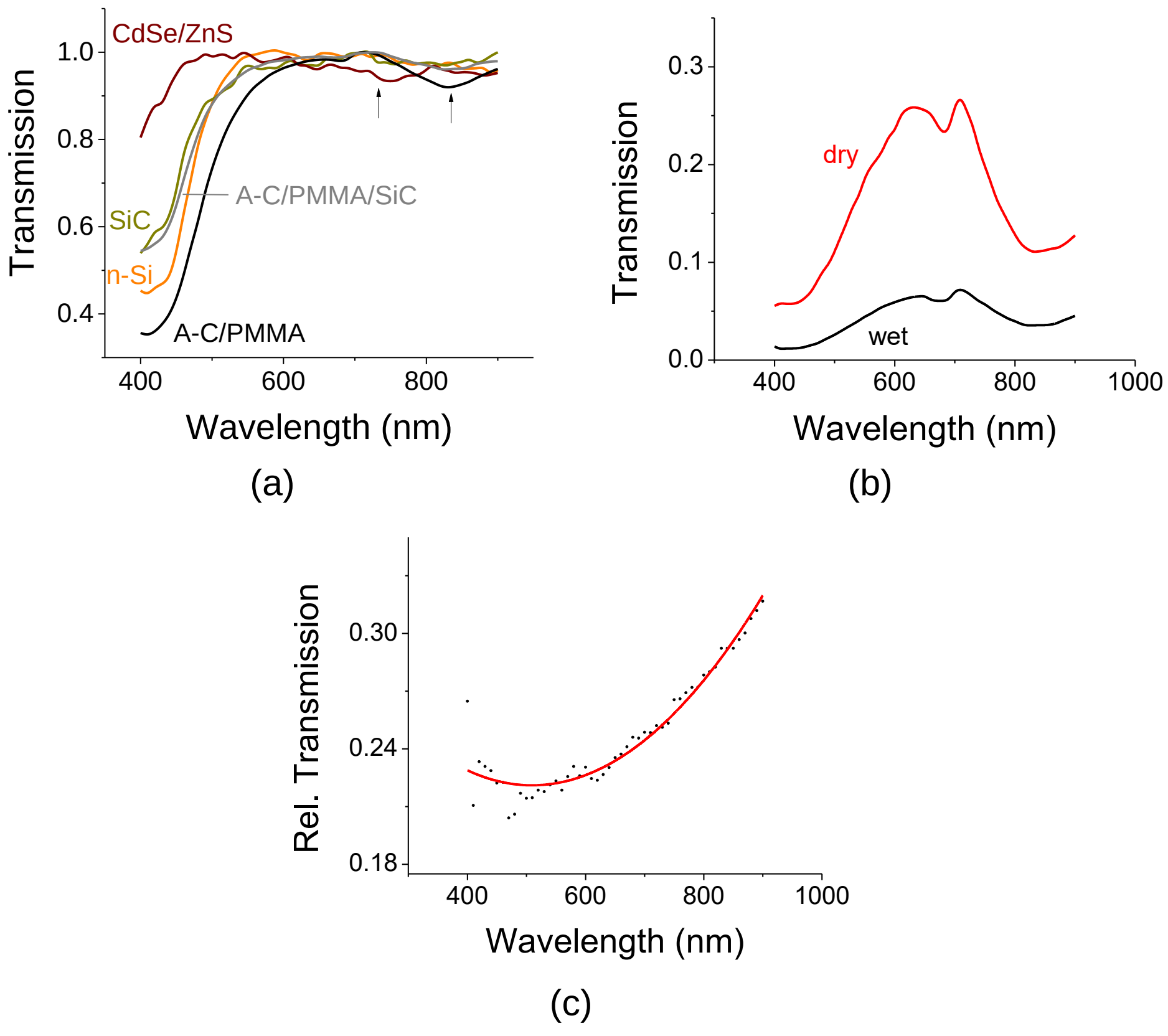

Fig. 2. Normalized optical transmission through the various film components used in the experiments. The peak transmission was set to 1 for each curve in order to accentuate the spectral response. The transmission of all curves were referenced to the signal through glass slide. The two arrows call attention to the absorption of CdSe/ZnS at ca 720 nm and the absorption at 840 nm, which is attributed to water.

One might assume that the relative absorption of the wet curve, when normalized by the dry curve would yield the absorption curve of merely the water. The absorption curve for water is minimal in the blue and maximal in the red. This is not the case here; the absorption trend is quite the opposite, namely, it is maximal in the blue/green and minimal in the red. The maximum absorption (the largest dip) is in the green wavelength range. This means that the wet/dry curve is mainly affected by scattering at the A-C solid/pore interface. Such scattering is susceptible to the refractive index of the pore filler (namely, water in our case). Of the two possible scattering mechanisms, Rayleigh and Mie scattering, one may favor the latter because of the small size of the pore and the $\pi$–$\pi^*$ bonds in the solid A-C. This plasmonic scattering is known for quantum carbon dots [13], and we propose that the scattering mechanism is similar to one occurs for inverse plasmonic structures.

***III.b. Optical Related Effects***

The capacitance change of SiC embedded, A-C based cell is described below. Capacitance increase can be observed in Fig. 3a,c. The relative capacitance increase, is normalized by the respective differences between the illuminated and non-illuminated areas: $\Delta C/C=[(C/A)_{illum}-(C/A)_{dark}]/(C/A)_{dark}$. Cyclic Voltammetry (CV) indicates a $\Delta C/C=27\%$ relative increase whereas the Charge Discharge measurements (CD) alludes to $\Delta C/C\sim 60\%$. Note that the current levels are larger when using CV. CV curves exhibit a tilt towards larger current values, as well as a broadening of the curve waist. A curve tilt without waist broadening is typically associated with sample heating as shown below. Fig. 3b shows the optical effect even in the absence of SiC particles; it amounts to $\Delta C/C\sim 14\%$ (including the effect of smaller exposed area) and is attributed to the A-C absorption in the blue. An asymmetric cell with two types of current collectors is shown in Fig. 3d. While the electrodes were both made of SiC embedded A-C, the current collectors were made of, respectively, ITO and Al. Illumination by the white light source was made through the front ITO film.

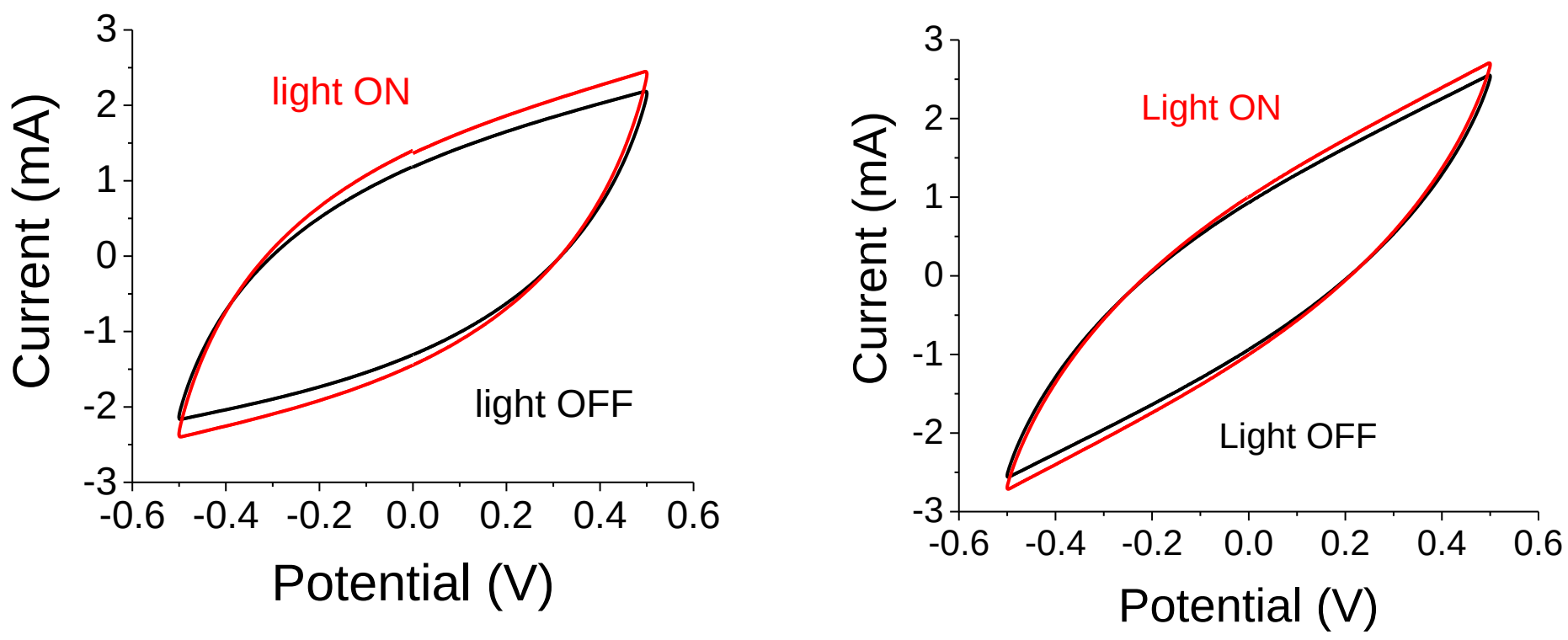

(a) (b)

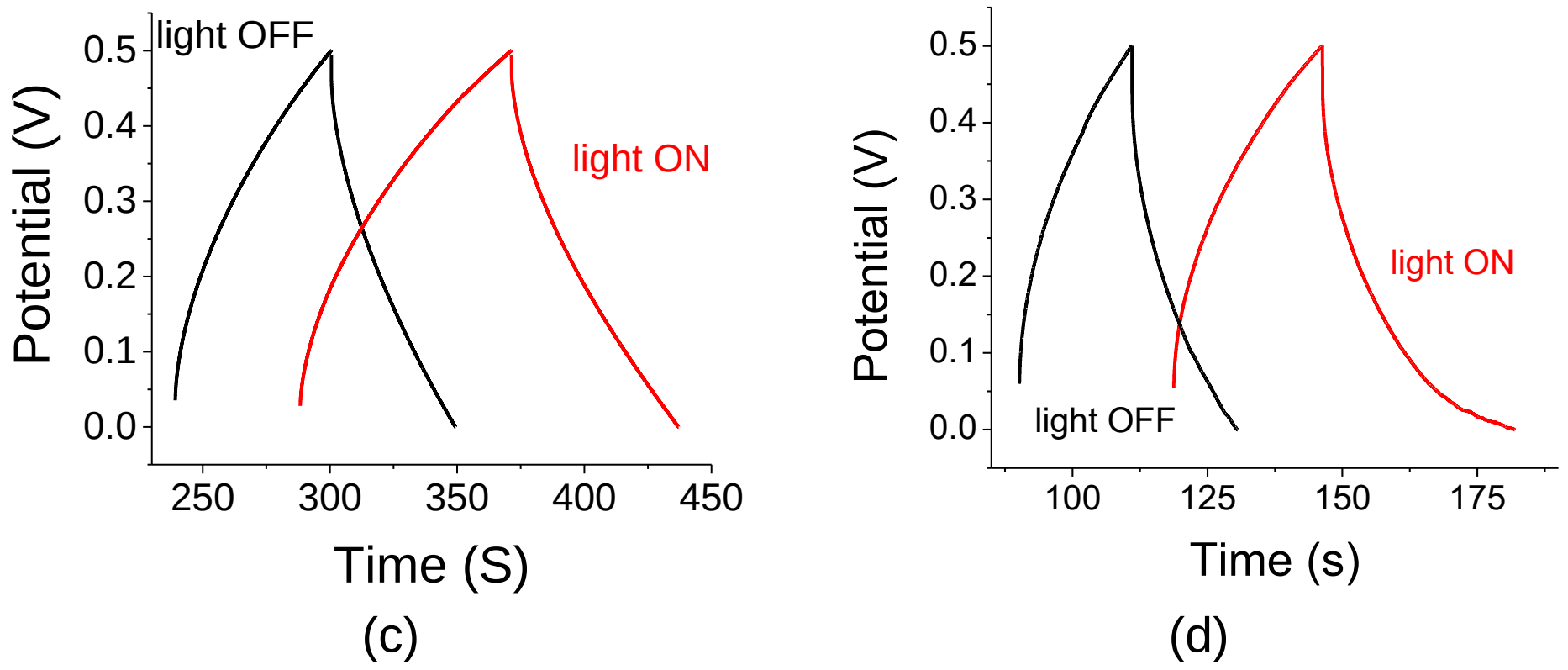

Fig. 3. (a) CV (at a rate of 0.1 V/s) when a sample is illuminated from the front (where the SiC particles are) and (b) when another sample is illuminated from the back (without SiC particles) (c) CD for sample (a) (excited with current of 0.2 mA) under light and dark conditions. (d) SiC embedded A-C electrode with ITO (front, facing the light) and Al (at the sample back) both used as current collectors.

The time response of the optically controlled capacitors is yet to be found. A first step is to assess the capacitance increase as a function of the illumination time. In Fig. 4 we show a capacitance increase of ca 50% for n-Si embedded, A-C electrode cell after 1 min and a further, smaller increase after 5 min of light exposure amounting to a total increase of 68%. Included in the assessment was the smaller exposure area of one half of the sample surface.

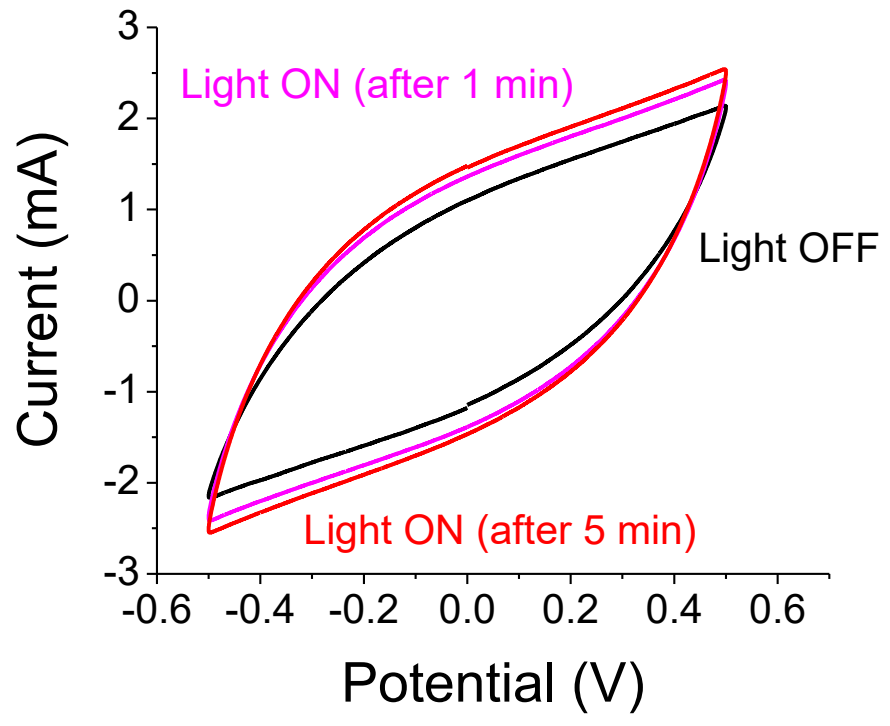

Fig. 4. Temporal response of n-Si embedded A-C based S-C. Most of the capacitance increase of 50% occur within the first minute of exposure while a fuller increase of 68% occurs after 5 minutes.

Further experiments were carried out using Electrochemical Impedance Spectroscopy (EIS) [14]. As shown in Fig. 5a,c, the white light illumination affects the electrode's impedance, an effect that amounted to 2%. A larger effect is noted for the (middle) diffusion region which becames more capacitive under illumination. Lastly, the differential capacitance exhibits a constant value in the frequency region between 50 to 125 mHz exhibiting a small increase under illumination (Fig. 5b,d). We note that a large double-layer capacitor exhibit smaller slopes in Fig. 5b,d.

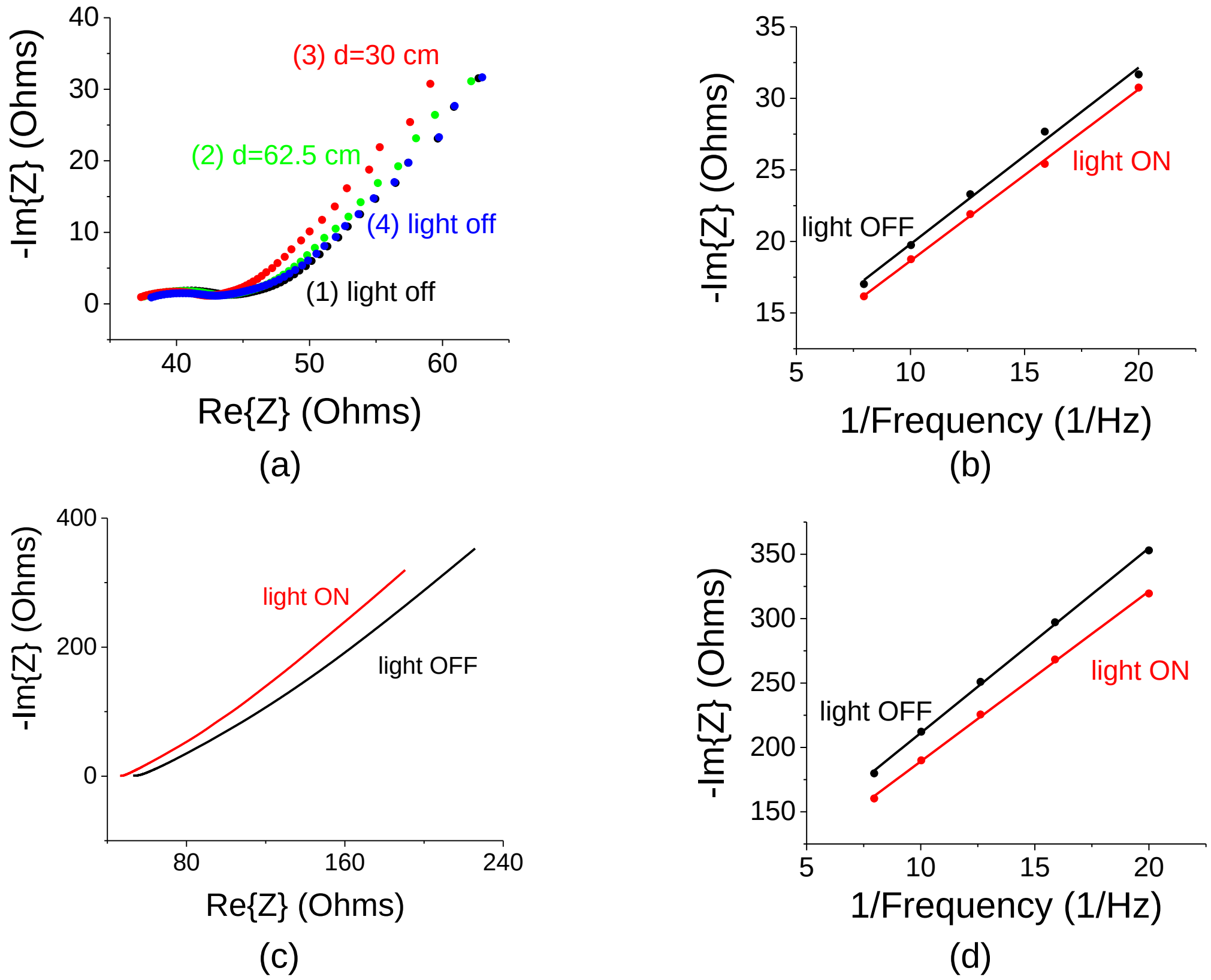

Fig. 5. (a) An EIS Nyquist curve as a function of light. (1) White light OFF. (2) White light ON placed at distance d=62.5 cm from the sample (translated to I=3 mW/cm$^2$ at the sample surface). (3) The white light was brought closer to the sample (d=30 cm, I=30 mW/cm$^2$). (4) The white light source was turned OFF and its related curve follows case (1). (b) Plot of –Im{Z} vs the inverse of the frequency exhibits a straight line whose slope is proportional to the differential capacitance. (c) A sample with a high (imaginary) impedance exhibits larger optical effect and its related differential capacitance curve is shown in (d). Note that under white light illumination, the slope of the curve, which is proportional to the inverse of the differential capacitance decreases

### *III.c. Thermal considerations*

The CV plot in Fig. 6a has been obtained when collecting the CV data while continuously heating the sample from 22 $^o$C to 36 $^o$C at the rate of 0.1 $^o$C/sec. Only the curves in the beginning and in the end of the scan are shown. The light was turned OFF. Note the slight rotation of the curve as the sample heats up; its waist at zero potential has changed a bit too. The relative change from the first scan at 22 $^o$C to the last at 36 $^o$C was <10%. Results for CD experiments indicate a larger thermal effect of <20% (Fig. 6b). Unlike CV experiments, here data were obtained with a hot plate at well stabilized temperatures of 24 and 36 $^o$C, respectively. Overall, the capacitance increase due to thermal effects are much larger than the thermal effect exhibited by n-Si embedded A-C [1]. The thermal effect with SiC particles may be attributed to the higher thermal conductivity of SiC in comparison with n-Si [15,16]. In Fig. 4c we show two CV curves for light OFF and light ON. Upon illumination, the temperature at the sample surface has elevated to 36 $^o$C. The 'as-is' relative capacitance change was 12% (without including the fact that the exposed area was half of the entire area of the S-C). It was 24% when normalizing by the smaller exposed area. On the other hand, while the trend (see also Fig. 3a) favors an optically related capacitance increase, one ought to acknowledge that the thermal and the optical effects are comparable and we cannot rule out the possibility of a heating effect by the optical source.

A very large thermal effect is obtained when the A-C electrodes are embedded with CdSe/ZnS quantum dots (Fig. 6d). The relative capacitance change due to the thermal effect is ca 35% compared to ca 10% when the white light is turned ON and OFF (not shown). While most samples (A-C embedded with SiC, or A-C embedded n-Si) reached ca 36 °C after a prolonged illumination, the unusual thermal effect in CdSe/ZnS embedded A-C electrode points to morphological changes, perhaps in the ligand that coats the dots. EIS for A-C embedded with CdSe/ZnS quantum dots exhibit a higher resistance at low frequencies rather than a higher capacitance trend (Fig. 6e)

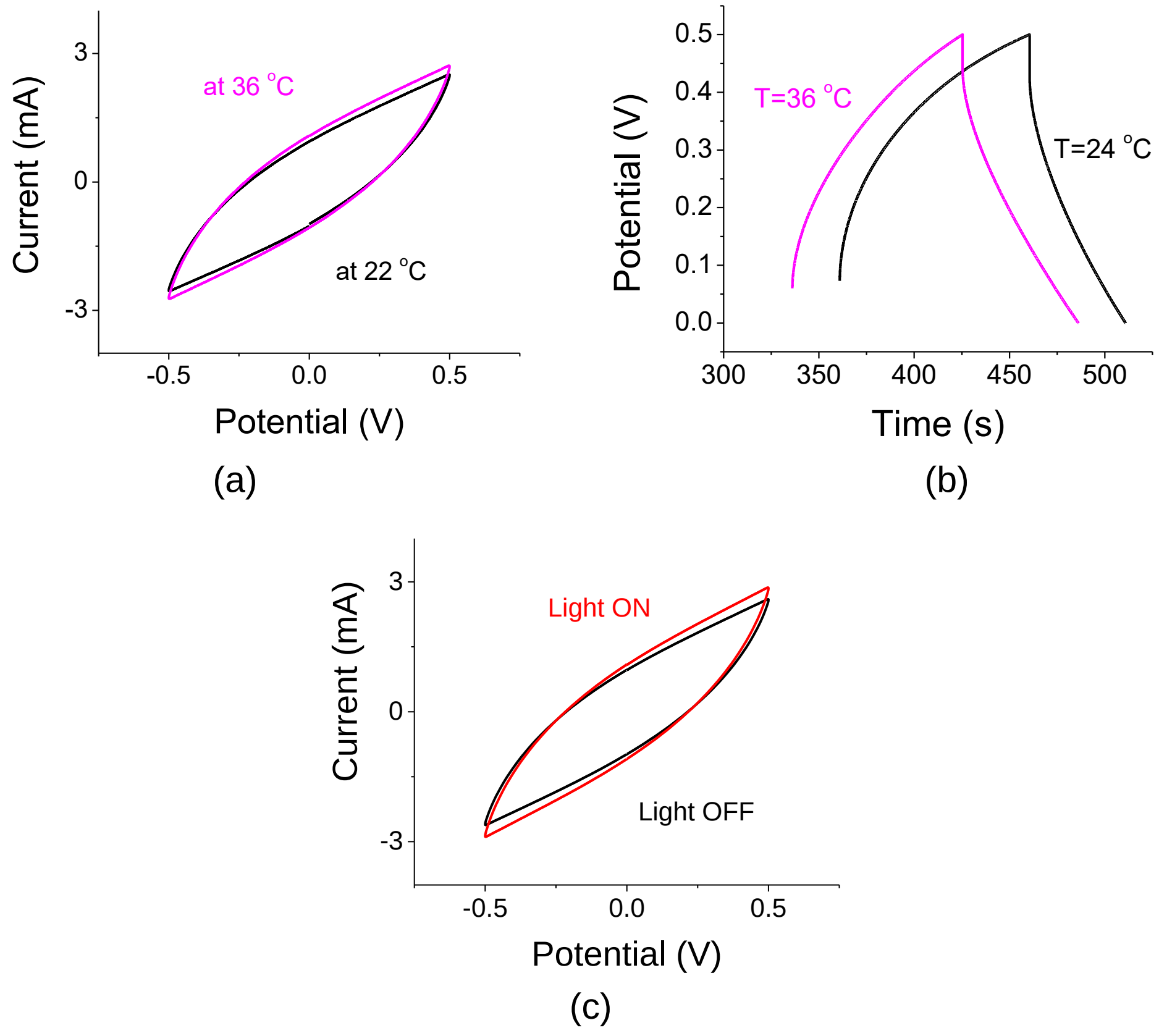

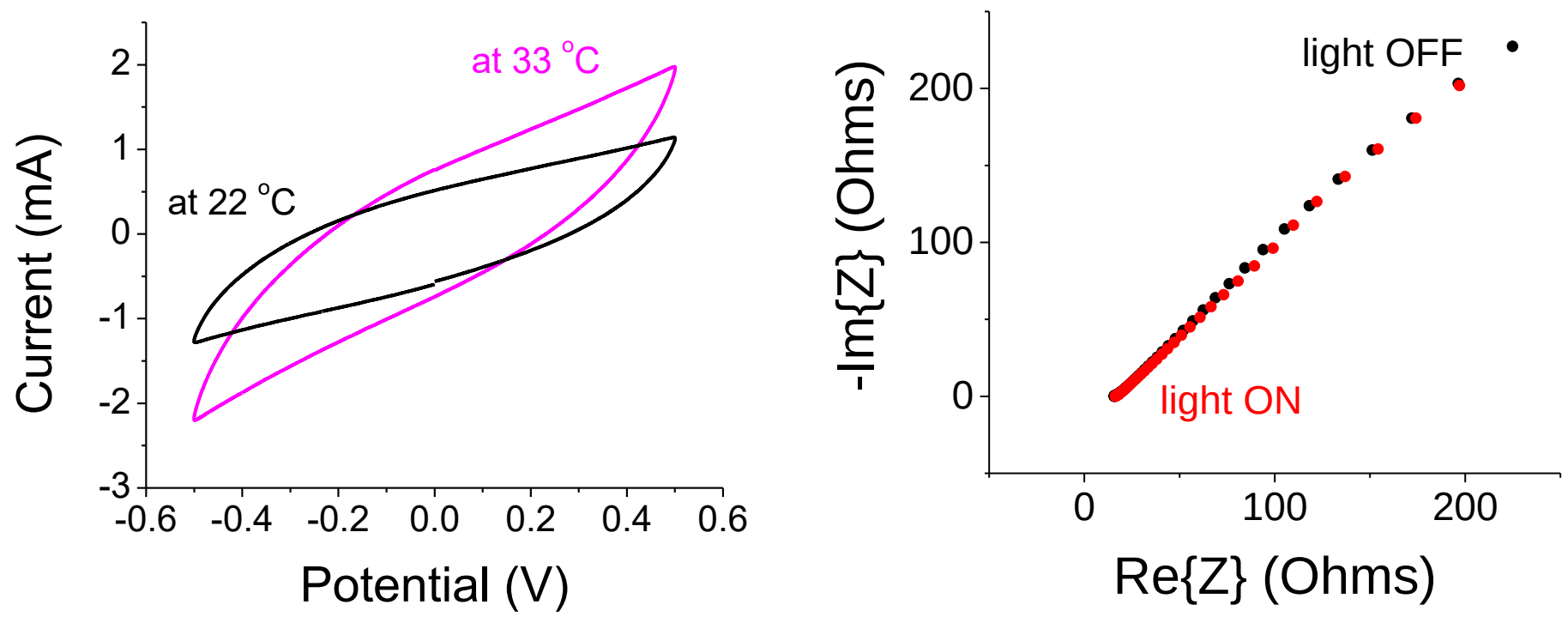

(d) (e)

Fig. 6. (a) The sample was heated when light was OFF from 22 to 36 °C. CV curves were obtained at a scan of 0.1 V/s. (b) CD curves when a sample was measured at two stabilized temperatures. (c) CV curves for light OFF (room lighting) and light ON. (d) CV for A-C electrode embedded with CdSe/ZnS quantum dots. The relative capacitance change due to the thermal effect is ca 35%. (e) EIS for A-C embedded with CdSe/ZnS quantum dots; the trend of the curve points to higher resistance at low frequencies.

### *III.d. Model*

A model was constructed to aid the understanding of these results. A small parallel plate capacitor filled with electrolyte was interfaced with two graphite-like electrodes. The electrodes were deposited on current collectors (terminals). The graphitic-like material was conductive; its dielectric constant was low, $\varepsilon_r$=2 and its conductivity was modest, $\sigma$=$10^5$ S/m. A charged particle (loosely defined here as QD) of radius 5 nm or 2.5 nm was placed inside the graphitic electrode. A CAD tool (Comsol) was used to determine the potential distribution and the increased capacitance when the terminals were biased by 1 V. The particle under illumination is modeled as either a dipole or as an ionized particle. In the dipole case, half the particle surface was decorated with one charge type and the other half was decorated with the opposite charge type. In the ionized particle case, the surface was decorated with only one charge type (either negative or positive).

The potential distribution in the absence of light induced charges is shown in Fig. 7a. Note that the potential is constant within the electrolytic layer (the green yellow region in the middle). The middle region is modeled as a floating potential region. Free charges are excited when light is absorbed by the semiconductor particles. These free charges are displaced by the external bias of the capacitor (±1 V in our case). The negative electrons are attracted to the positive terminal, thereby attracting more positive charges to the terminal from the biasing source. Similarly, positive induced charges are attracted to the negative terminal. Such a self-promoting process results in an overall increase in the cell's capacitance, effect that was used to increase the dielectric constant of materials, and its concept is known as artificial dielectrics, AD [12-13]. The AD concept has originally been devised for the optical and microwave frequencies. Here we adopt it to very low frequencies and direct current (DC) cases. The introduction of a dipole at the electrode, thus alter the potential distribution at the terminal (Fig. 7b). Since the terminal voltage is kept constant, the cell capacitance is related to the charge that is accumulated on the terminal When the light intensity increases, so will the induced charge on the particle surface and the cell capacitance (Figs. 7c,d). In Figs. 7c,d, a negative QD surface charge value means that the dipole is not made of free carriers but is frozen for some reason; its negatively charge hemisphere is pointing away from a positively biased terminal.

A particle is ionized when electrons are donated to the A-C electrode. The remaining positive charge increases the capacitance if the nearby terminal is negatively biased and decreases the cell capacitance if this terminal is positively biased. (Fig. 7e). An ionized particle, whose some of its electrons have been stripped away, would impact the cell capacitance in an asymmetrical fashion (which we do not see in the experimental data).

As a note, a surface charge of 5 $mC/m^2$ translates to ca 6.25 and 2.5 electrons per particle of radii r=5 and r=2.5 nm, respectively.

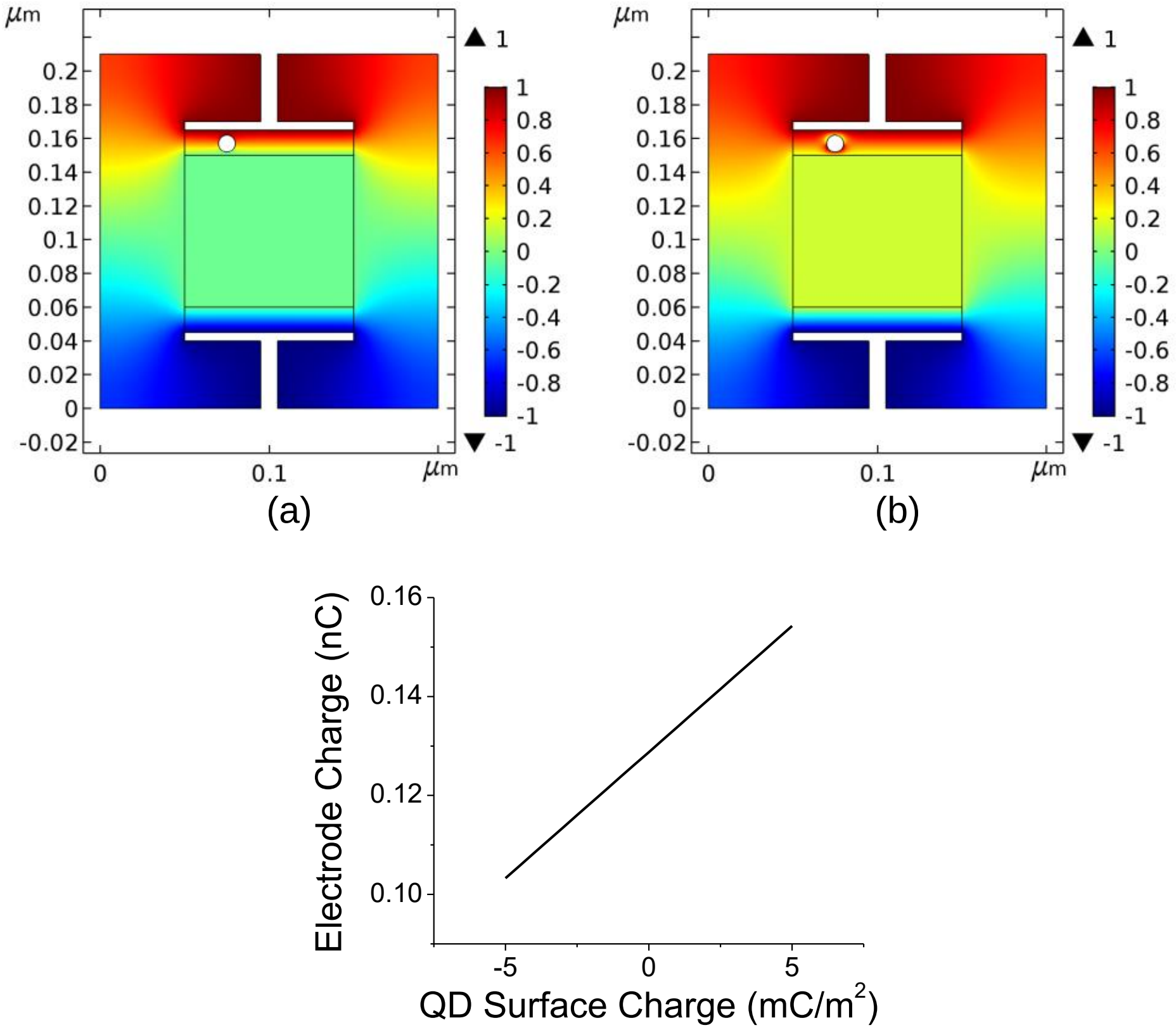

(c)

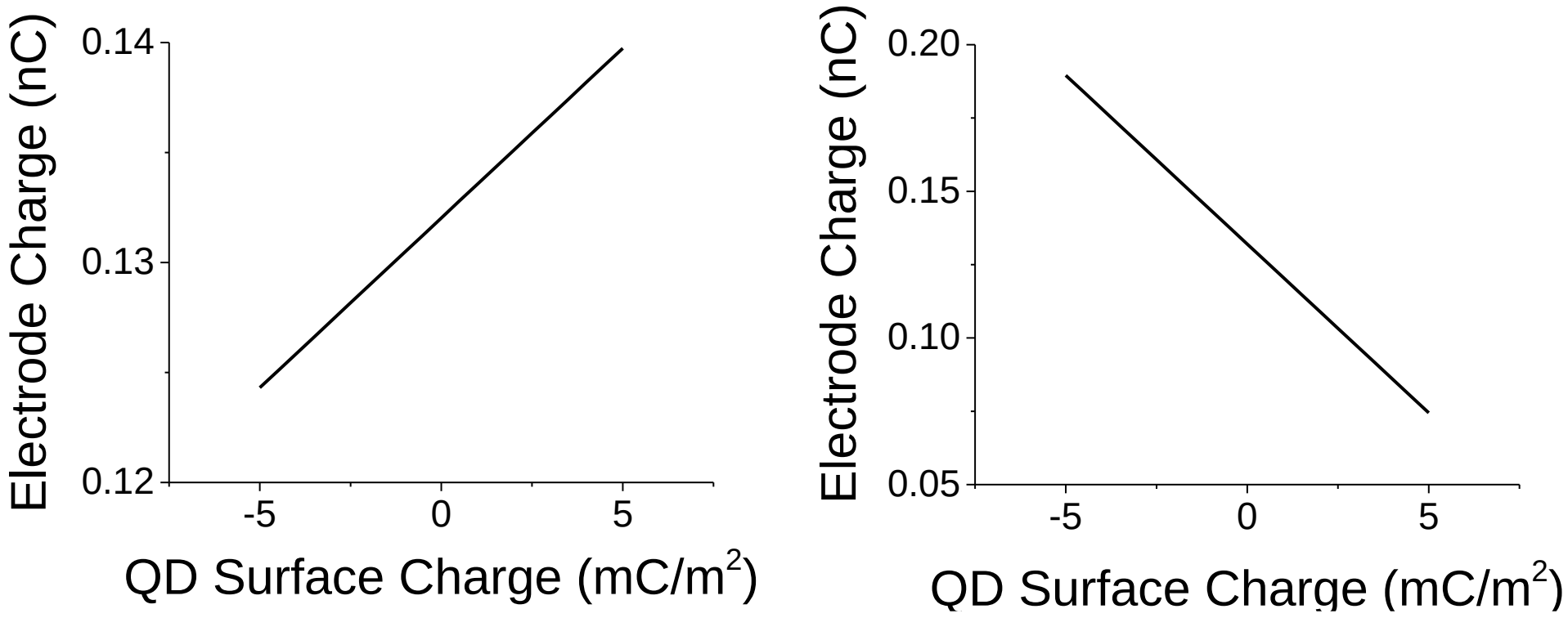

(d) (e)

Fig. 7. (a) Potential distribution when uncharged QD particle is embedded in a graphitic-like electrode. The cell is biased by ±1 V. (b) Potential distribution when the 5 nm particle acts like a dipole. (c) Change in the cell capacitance while varying the surface charge density of the dipole for an r=5 nm particle. The surface charge type is (d) Change in the cell capacitance while varying the surface charge density for an r=2.5 nm dipole. Note the relatively large effect even though the charge has reduced by a factor of 2.5. (e) Change in the cell capacitance while varying the surface charge density of an ionized QD particle with r=2.5 nm.

**Conclusions**

Supercapacitors whose A-C electrodes were embedded with submicron SiC particles have shown a capacitance increase that is attributed to optically-induced local dipole effect. The effect though is smaller than for n-Si embedded electrodes. The light absorption by the dark A-C also leads to capacitance increase through thermal effects. The thermal effect has been shown to be larger than the related effect for n-Si embedded electrodes. Finally, a large anomaly is reported when CdSe/ZnS QD are embedded in A-C electrodes; there the thermal effect is much larger than the optical one. Further understanding of these phenomena may help us design better optically controlled S-C elements.

**Acknowledgement**

To Dr. X. Zhang of NJIT for SEM and AMF pictures.

1. H. Grebel, C 2021, 7(1), 7; https://doi.org/10.3390/c70100072.
2. Keh-Chyun Tsay, Lei Zhang, Jiujun Zhang, Electrochimica Acta 60 (2012) 428–436.
3. Yudong Li, Xianzhu Xu, Yanzhen He, Yanqiu Jiang and Kaifeng Lin, Polymers 2017, 9, 2; doi:10.3390/polym9010002.
4. M. Kaempgen, C. K. Chan, J. Ma, Y. Cui, and G. Gruner, Nano Letts., , 9 (2009) 1872.
5. Michio Inagaki, Hidetaka Konno, Osamu Tanaike, Journal of Power Sources 195 (2010) 7880–7903.

6. Zhang, S and Pan, N, 2015. DOI:10.1002/aenm.201401401. https://escholarship.org/uc/item/26r5w8nc

7 Jeffrey W. Long, Daniel Bélanger, Thierry Brousse, Wataru Sugimoto, Megan B. Sassin, and Olivier Crosnier, MRS BULLETIN vol. 36, 2011.

8. Yuanlong Shao, Maher F. El-Kady, Jingyu Sun, Yaogang Li, Qinghong Zhang, Meifang Zhu, Hongzhi Wang, Bruce Dunn, and Richard B. Kaner, Chem. Rev. 2018, 118, 9233−9280. DOI: 10.1021/acs.chemrev.8b00252.

9. Mohammad S. Rahmanifar, Maryam Hemmati, Abolhassan Noori, Maher F. El-Kady, Mir F. Mousavi, Richard B. Kaner, Materials Today Energy 12 (2019) 26-36. https://doi.org/10.1016/j.mtener.2018.12.006

10. Xin Miao, Roberto Rojas-Cessa, Ahmed Mohamed and Haim Grebel, IEEE IoT, 2018

11. Roberto Rojas-Cessa, Haim Grebel, Zhengqi Jiang, Camila Fukuda, Henrique Pita, Tazima S. Chowdhury, Ziqian Dong and Yu Wan, Environmental Progress & Sustainable Energy, 37, (2018) 155-164. DOI 10.1002/ep.

12. Emre O. Polat and Coskun Kocabas, Nano Lett. 2013, 13, 5851−5857. dx.doi.org/10.1021/nl402616t.

13. Indrajit Srivastava, John S. Khamo, Subhendu Pandit, Parinaz Fathi, Xuedong Huang, Anleen Cao, Richard T. Haasch, Shuming Nie, Kai Zhang, and Dipanjan Pan, Adv. Funct. Mater. 2019, 29, 1902466. DOI: 10.1002/adfm.201902466

14. Bing-Ang Mei, Obaidallah Munteshari, Jonathan Lau, Bruce Dunn, and Laurent Pilon, J. Phys. Chem. C 2018, 122, 194−206. DOI: 10.1021/acs.jpcc.7b10582.

15. http://www.ioffe.ru/SVA/NSM/Semicond/SiC/thermal.html

16. Zhaojie Wang, Joseph E. Alaniz, Wanyoung Jang, Javier E. Garay and Chris Dames, Nano Lett. 2011, 11, 2206–2213. dx.doi.org/10.1021/nl1045395